# Metadata and data management for the Keck Observatory Archive

H. D. Tran*[a], J. Holt[a], R. W. Goodrich[a], J. A. Mader[a], M. Swain[b], A. C. Laity[b], M. Kong[b], C.R. Gelino[b], G. B. Berriman[b]

[a]W. M. Keck Observatory, 65-1120 Mamalahoa Hwy., Kamuela, HI, USA 96743; [b]NASA Exoplanet Science Institute, Mail Code 100-22, 770 South Wilson Ave., Pasadena, CA, USA 91125

## ABSTRACT

A collaboration between the W. M. Keck Observatory (WMKO) in Hawaii and the NASA Exoplanet Science Institute (NExScI) in California, the Keck Observatory Archive (KOA) was commissioned in 2004 to archive observing data from WMKO, which operates two classically scheduled 10 m ground-based telescopes. The observing data from Keck is not suitable for direct ingestion into the archive since the metadata contained in the original FITS headers lack the information necessary for proper archiving. Coupled with different standards among instrument builders and the heterogeneous nature of the data inherent in classical observing, in which observers have complete control of the instruments and their observations, the data pose a number of technical challenges for KOA. For example, it is often difficult to determine if an observation is a science target, a sky frame, or a sky flat. It is also necessary to assign the data to the correct owners and observing programs, which can be a challenge for time-domain and target-of-opportunity observations, or on split nights, during which two or more principle investigators share a given night. In addition, having uniform and adequate calibrations are important for the proper reduction of data. Therefore, KOA needs to distinguish science files from calibration files, identify the type of calibrations available, and associate the appropriate calibration files with each science frame. We describe the methodologies and tools that we have developed to successfully address these difficulties, adding content to the FITS headers and "retrofitting" the metadata in order to support archiving Keck data, especially those obtained before the archive was designed. With the expertise gained from having successfully archived observations taken with all eight currently active instruments at WMKO, we have developed lessons learned from handling this complex array of heterogeneous metadata that help ensure a smooth ingestion of data not only for current but also future instruments, as well as a better experience for the archive user.

## 1. INTRODUCTION

The Keck Observatory Archive (KOA)[1,2,†] began operations in August 2004, initially serving level-0 (uncalibrated) data from the High-Resolution Echelle Spectrometer (HIRES). It now serves data from all eight active WMKO facility instruments. Also available are some level-1 data, namely extracted spectra from HIRES, calibrated images from the Near-Infrared Camera (NIRC2), and reduced data cubes from the integral field unit OH-Suppressing Infra-Red Imaging Spectrograph (OSIRIS). The latter two instruments are used behind the adaptive optics (AO) system.

Since KOA was developed well after WMKO began operations in 1993, much of the data produced were not designed with modern archiving in mind or intended to be released publicly. As a result, the data as generated from the instruments are not fit to be ingested directly into the archive because they lack the information necessary for appropriate identification, cataloging, and later querying and retrieval. Such information is essential to ensure the correct assignment of data to the PIs (principal investigators), their proper protection, and their eventual release to the public after a proprietary period has passed. The data evaluation and preparation (DEP)[3] process performed at WMKO makes these data suitable for archiving by adding additional keywords, or metadata, to the FITS headers. These metadata include a unique identifier for each file, the PI and program information, the status of the instrument, some image statistics for quality assessment, and the type of data – whether it is a science object or calibration file and the type of calibration.

Determining the PI and type of data for a file can sometimes be a challenge since WMKO uses the classical observing mode, in which the observers have complete control of the observations and what they input for certain keywords in the FITS header. Unlike service or queue observing, in which the observations are highly structured and meticulously planned, observers at WMKO are afforded a high level of flexibility, and not required to follow any strict standards or protocols. In addition, many keywords that would have helped to determine the file data type simply did not exist. Due to the different choices by different instrument builders, the metadata structure is also not uniform across all instruments. As a result, the information contained in the native headers is heterogeneous and can often present a challenge in determining the pedigree of the data files.

---

*htran@keck.hawaii.edu
†http://koa.ipac.caltech.edu/

Because of this highly flexible, customer-service environment that aims to provide a user-friendly observing experience at Keck, KOA has had to be very vigilant in anticipating the needs and intentions of the observers so that all data can be properly archived. We will describe some software techniques that we have used to determine the appropriate information that can be used to retrofit the metadata to support archiving, especially legacy data, those obtained in the past before the archive began operations, and those from split nights, in which more than one PI and program are assigned for a given night. We offer lessons learned from our experience in managing these sets of heterogeneous data from WMKO.

## 2. PROCESSING LEVEL-0 DATA

### 2.1 Image typing

Due to W. M. Keck Observatory's classical observing mode, different instrument designers, and lack of original intent to make data publicly available through an archive, each instrument has slightly different and sometimes non-standard set of FITS header keywords. Moreover, the type of data, i.e., on-sky science target or calibration, is not always explicitly recorded in the FITS headers. The first step in data processing by KOA not only ensures that the necessary keywords and metadata exist and are in the correct format, but also determines the type of data for each file, a process we call "image typing", which is custom to each instrument, following a method described in Tran et al. (2012). We presented methods for image typing NIRC2 images in that paper. Here we describe the technique for image-typing OSIRIS data (Fig. 1).

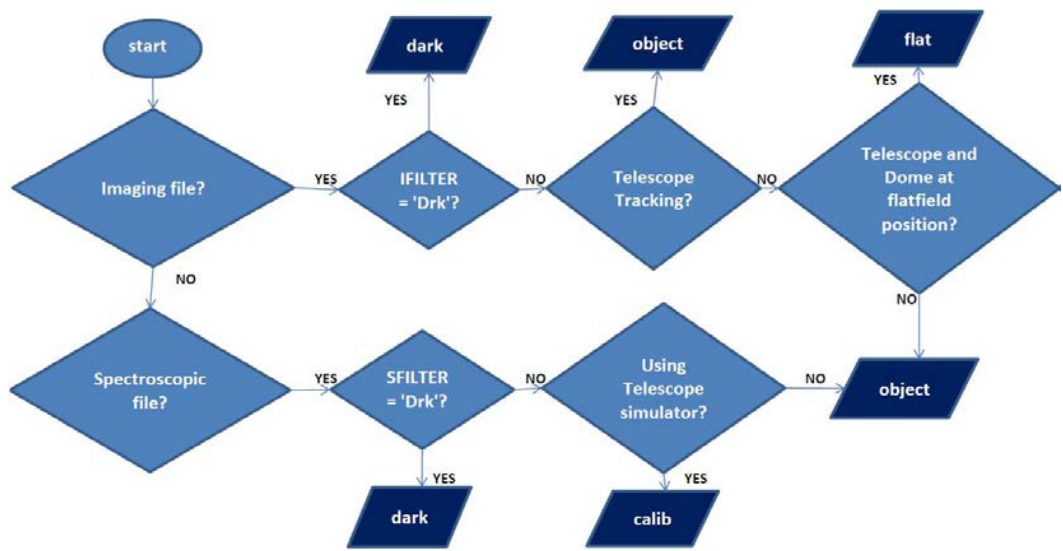

Figure 1. Flowchart for determining image types of OSIRIS data.

### 2.2 World coordinate system (WCS) keywords

In general, the imaging data obtained at Keck do not contain proper world coordinate system (WCS) information. The rare exception is data from the OSIRIS instrument, which was also delivered with a data reduction pipeline (DRP), due to the complexity of the data. Since having WCS would be useful and highly desirable, as part of preparing NIRC2 data for archiving, WCS keywords have been added to the image FITS header. The NIRC2 WCS information is determined from the keywords RA, DEC, ROTPOSN, using an instrument PA zero-point of 0.7° (keyword INSTANGL) with a correction of -0.252° for the narrow camera[4], and the pixel scales in the FITS header. In addition, for vertical angle and stationary rotator modes, the parallactic angle (PARANTEL or PARANG) and elevation (EL) are also used. In "vertical angle mode", the field is maintained at a fixed angle relative to the vertical (elevation) axis, while in "stationary mode", the rotator is not tracking but remains stationary at a fixed rotator angle.

For other instruments, we plan to add WCS keywords to the imaging data from LRIS, DEIMOS, and ESI when the full set of keywords is ingested in the future. Currently, only HIRES, NIRSPEC and NIRC2 data in the archive are ingested with the full sets of keywords. For the rest of the instruments, only a limited set of meta-keywords representing the most commonly used ones, is available. Only the keywords most commonly queried by archive users or needed by the archive itself are cataloged. This was adopted as a strategy to support a quicker release of more data from more instruments.

**2.3 Header keyword oddities**

The heterogeneous nature of the headers in Keck data manifests itself in several ways, two examples of which are missing and duplicated keywords. We describe here briefly the nature of these problems and how KOA deals with them.

**Missing keywords:**

Occasionally, for various reasons the data FITS headers are found to be missing some critical keywords required for data ingestion, such as UTC, RA, DEC, and DATE-OBS. In general, KOA does its best to recover such keywords so that as much of the data can be archived as possible, but a small fraction of the data remain unrecoverable. If so, the data are rejected by KOA, and the archive user will never see them. For some data, DATE-OBS and UT can be recreated by using the date and time stamps of the original data directory and files as they were written to disk. For others, like NIRSPEC calibrations and OSIRIS darks, such information was lost and another method is used, as described below.

As delivered, the script to take darks for OSIRIS "osirisTakeDarks" turned off writing the telescope "drive control system" (DCS) and AO keywords to the headers. It was probably believed by the instrument designers that since these were dark frames taken during the day, it was not essential to know the status of the telescope or AO system. One consequence of not writing DCS keywords is that headers contain no time information: missing is the crucial UTC keyword, since it comes from the telescope system. In order to ingest these files into the archive, we had to add fictitious UT time stamps into these dark calibration files beginning with UT midnight of the day the files were taken, and then using the exposure times of the dark frames to separate them.

For a number of NIRSPEC afternoon calibrations, the first UTC that must be recreated is given a value of 00:00:00. Subsequent files are assigned times in 5 second increments. Files taken during the night and having missing UTC values are assigned a UTC approximately midway between the UTC times of adjacent files which have valid times.

All reconstructed keyword values can be identified by the comment line in the FITS header. Recreated values of UTC or DATE-OBS have the comment: "Original value missing – added by KOA".

**Duplicated UTC keywords:**

NIRC data contain two duplicated UTC keywords with slightly different values, marking the beginning and end of an exposure. In this case, using NASA's CFITSIO utility "modhead", we read and ingest only the first instance of UTC.

**"Unknown", "null", invalid values or format in keywords:**

The software server running NIRSPEC occasionally crashes during observing, causing some motor keywords to take on the value of "unknown". These keywords can be restored to their proper values with a server recovery, but only after an initialization of the motors. However, sometimes it is desirable not to initialize the motors immediately following a server recovery, in order to preserve the instrumental settings, leading these keywords to retain the "unknown" value. In such cases, KOA populates these keywords with values just prior to the crash.

RA and DEC are also found to take on the value of "*null*" for calibration frames taken during the day, when DCS is unavailable at times. For these cases, we simple accept "*null*" upon ingestion.

On occasion, keyword values or formats are modified or added without proper documentation, causing them to be out of range and failing the keyword validation step in our ingestion process. The accepted range or list of values for these keywords is adjusted on a case by case basis as they arise. An example is the DATE-OBS keyword in NIRC data. Some of the early data took on the form "DD/MM/YY" instead of the usual standard format "YYYY-MM-DD". DATE-OBS for these data was changed to the correct format before they were ingested.

# 3. TOOLS TO MANAGE METADATA

## 3.1 Observer's tools: PIG and CAT

The metadata (FITS headers) for each file is validated before ingestion into the archive. If needed, additional keywords are added to support archival use. To facilitate the metadata retrofitting process, we have provided observers at WMKO with certain tools to be used while observing. We've developed an algorithm[3] to assign the correct PI and program for the data file, which is especially useful for legacy data. For data currently obtained at the telescopes, a tool called "program identification GUI" (PIG) can be used to set certain keywords (i.e., OBSERVER for the observers' names, and OUTDIR for the output data directory) in the header that will facilitate the assignment of data to the proper owners. This tool was formerly referred to as the "split-night GUI" and briefly described in Tran et al. (2012), but an additional challenge is to identify the true owners of data on nights when the program is not scheduled, such as "targets of opportunity" (ToOs) programs (for example, a nearby supernova, or interesting activity in the Galactic Center). We have since redesigned the PIG to support these programs. ToO or time-domain astronomy programs are not regularly scheduled, but instead will be called in to interrupt the scheduled observing for some relatively quick observations of the transient phenomenon. With a list of approved ToO programs from the time allocation committees, part of the GUI expands to allow selection of one of these programs should an interrupt be triggered. Again, this will easily and clearly allow ToO data to be assigned to the proper PI, and properly separated into a new directory. The PIG logs all actions from interface input to monitor changes in the above keywords. These logs can then be parsed for data assignment.

The redesigned PIG with an example of the expanded list of ToO programs is shown in Figure 2. The "info" button for each ToO program may contain information on special conditions or requirements that allow the ToO to be triggered, such as target coordinates, environmental conditions, and instruments.

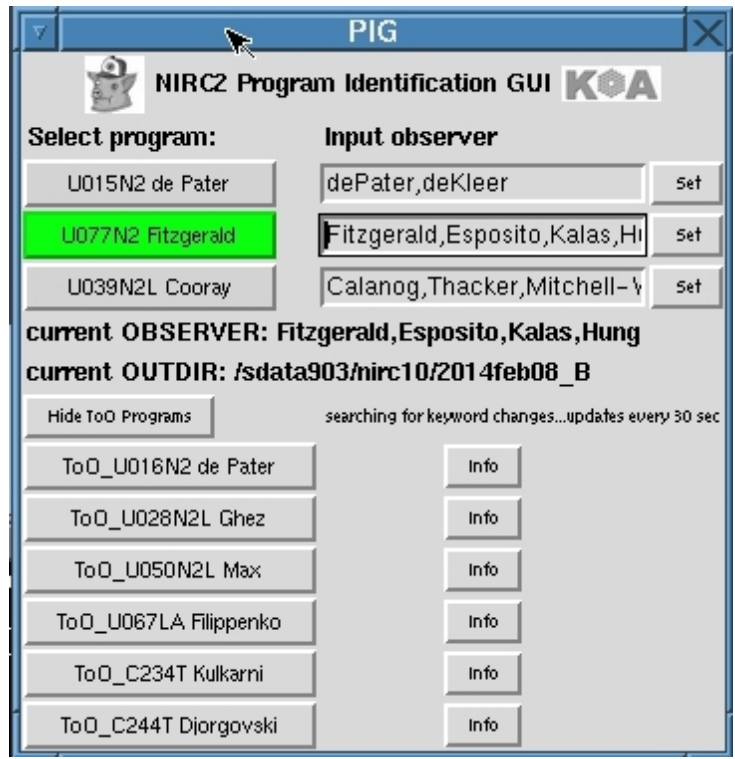

Figure 2. The program identification GUI (PIG) with support for ToO programs. The active program is indicated in green. To switch programs, observers click on the appropriate button, which sets the OBSERVER and OUTDIR keywords.

Another tool that KOA has developed to help observers is the "calibration acquisition tool" (CAT) for many of the instruments. We encourage observers to adopt a more structured approach by making it easier for them to take calibration data. The CAT is a simple interface that allows observers to easily select and acquire the necessary calibrations for the science data all at once. With one press of a button, observers can obtain a sequence of calibrations, rather than manually reconfiguring the instrument for each type of calibration. An example is shown in Figure 3 for HIRES. Observers simply select the types of calibrations, the desired exposure time for each, and the quantities for each type and click "Go". While the sequence is running, they can concentrate on other tasks or go eat dinner. An option to shutdown the instrument after the calibration acquisitions are finished is available, so that observers can retire to bed after their observing sessions in the morning without having to wait for the sequence to complete. For NIRSPEC, it may not even be necessary to input the exposure time, as the CAT automatically calculates the appropriate exposure times based on the instrumental settings (i.e., filters, spectral resolutions, etc…). This tool should help ensure that the necessary calibrations for the science files are available and uniformly obtained.

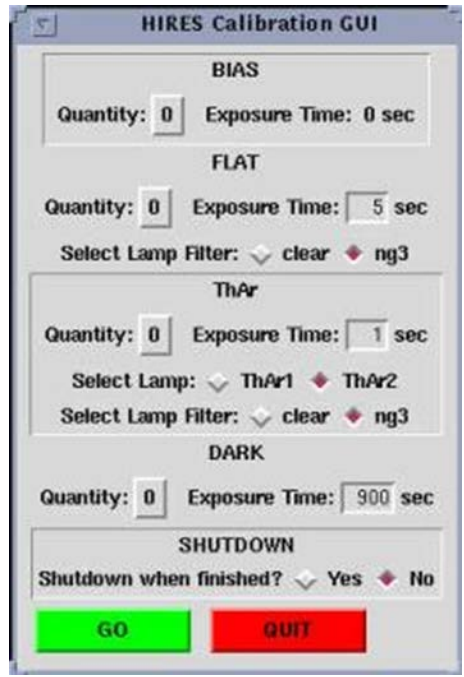

Figure 3. The calibration acquisition tool (CAT) for HIRES.

### 3.2 File names and translator

The names of the data files produced from WMKO instruments do not follow any strict standards. In fact, flexibility has been preferred over structure. Usually a default prefix, for example "hires", is given, followed by a frame number, which together makes up the name of the FITS file. Thus, for example "hires3103.fits" is a file name coming out of HIRES, or "n0012.fits" is one typically expected for a NIRC2 file. In addition, observers are free to change this prefix at any time. This can lead to ambiguity with regard to the instrument or date of the file. To support ingestion of data from all instruments, it was necessary for KOA to standardize the names of the data files. These unique identifiers, called KOAIDs, consist of a two letter code representing the instrument, the UT date of observations (in the form YYYYMMDD), and the number of seconds after UT midnight. For example, the files named "HI.19960727.55403.fits", "MF.20121004.23746.fits", and "OS.20070407.23368.fits" represent a HIRES, MOSFIRE, and OSIRIS frame taken in 1996 July 27 UT, 2012 October 04 UT, and 2007 April 27 UT, respectively.

However, these KOAIDs are unfamiliar to WMKO PIs and observers, and not friendly to data reduction tools that may expect the original names of the files as they come out of the instruments. This represents the main reason that discourages observers (and PIs) from using KOA to download their data. KOA provides a UNIX script that uses FITS keywords in the header to translate the KOA filenames back to the original filenames that users are familiar with. This should encourage more observers to use KOA to access their data.

# 4. LEVEL-1 DATA PROCESSING

KOA currently offers level-1, or extracted, reduced or calibrated data products, for three instruments: HIRES, NIRC2 and OSIRIS. These level-1 data are produced by automated pipelines and intended for browsing purposes only, and may not be of publishable quality. In particular, in the case of near-infrared instruments, no data stacking of dithered images (or sky subtraction, in the case of NIRC2) has been performed. For high-quality scientific use, users are recommended to download and reduce the raw data themselves.

## 4.1 Calibration association

One aspect of KOA that's useful but potentially challenging for a ground-based archive is the matching of appropriate calibration files with the science files. It is important for users to be able to find the appropriate calibrations so that the science data can be properly calibrated and reduced. In order to achieve this process, which we refer to as "calibration association", it is necessary to first identify and categorize the data into image types, as described in Tran et al. (2012). The image type of the file is then written in the header as a meta-keyword KOAIMTYP. Once the image typing is done, we can follow a set of procedures to associate calibration files with the science files. The calibration association as described below is available only for those instruments with the full set of ingested keywords.

In general, a calibration file is associated with a science observation by comparing those keywords that define both the instrument and detector (CCD) configurations. A pair of science and calibration files is said to be associated if these keyword values match, within the specified tolerances (see Table 1 for an example for HIRES). For bias and darks, only the CCD configuration keywords must match; for the rest of the calibration file types, both the instrument and CCD configurations must match.

By default, KOA returns all calibration files associated with the science observation on the night of the observation. If fewer than 5 flats, 1 bias, 1 dark or 1 arclamp are returned, the user interface will continue searching on adjacent nights until at least these numbers of each type are found for the most stable instruments, such as HIRES and NIRC2. For some types of calibrations, such as "trace" and "focus", it will not search beyond the original night. Specifically, it searches in ± one day intervals, up to ± 3 days on either side of the science observation. If the minimum number of each calibration file type is not found in the ± 3 day interval, the user interface will return whatever matching calibration files are available in that period. For some instruments, like NIRSPEC, whose calibrations display measurable variations from night to night, the calibration association is restricted to files taken on the same day of observations.

Table 1. Instrument and CCD configuration keywords to match for the calibration association of HIRES data

| INSTRUMENT CONFIGURATION KEYWORDS | DESCRIPTION | UNIT | DATA TYPE | EXAMPLES | TOLERANCE |
|---|---|---|---|---|---|
| SLITWID | Slit width | mm | float | 10.21383190 | ±0.005 |
| SLITWIDT | Slit width projected on sky | arcsec | float | 0.861 | ±0.005 |
| FIL1NAME | Filter #1 | - | char | KV370 | |
| FIL2NAME | Filter #2 | - | char | clear | |
| COLLRED | Red collimator in? | - | char | T/F | |
| COLLBLUE | Blue collimator in? | - | char | T/F | |
| ECHANGL | Echelle angle | deg | float | 0.01441866 | ±0.001 |
| XDISPERS | Cross disperser type | - | char | RED | |
| XDANGL | Cross disperser angle | deg | float | 0.32800001 | ±0.001 |
| CCD CONFIGURATION KEYWORDS | | | | | |
| AMPMODE | Amplifier mode | - | char | SINGLE:B | |
| AMPLIST | Amplifier list | - | char | 3,1,0,0 | |
| BINNING | Binning | pixels | char | 2,1 | |
| MOSMODE | Mosaic readout mode | - | char | B, G, R | |
| NVIDINP | No. of video inputs | - | int | 3 | |
| PRECOL | Pre-column pixels | pixels | int | 12 | ±0 |
| POSTPIX | Post-column pixels | pixels | int | 80 | ±0 |
| CCDGAIN | CCD gain setting | - | char | low | |

**4.2 Automating the OSIRIS data reduction pipelines (DRPs)**

A somewhat related aspect of finding the appropriate calibration files is identifying the correct sky frame for background subtraction, which is important for near-infrared data. This is important if the DRP is to proceed automatically without manual intervention. The DRP delivered with OSIRIS is highly sophisticated, and has an online mode for quick real-time reductions which are used primarily for basic data visualization and quality assessment done at the telescope while observing. The offline version of the data reduction system includes an expanded reduction method list, does more iterations for a better construction of the data cubes, and is used to produce publication-quality products. The most important step in fully automating the DRP for OSIRIS data reduction is the identification of the sky frame for background subtraction. This process has been described in Holt et al. (2014), which can be consulted for more details. The flowchart for auto-identification of the sky frame is outlined in Figure 4.

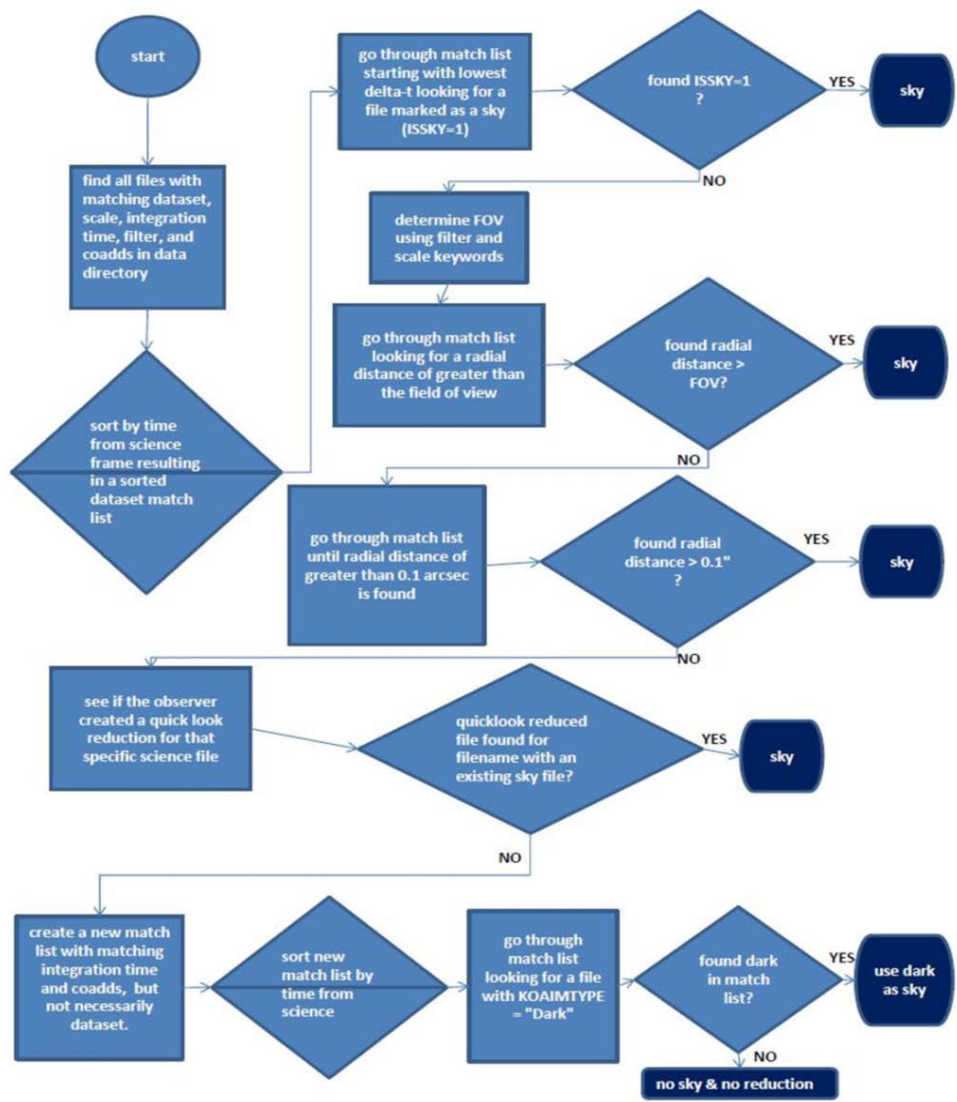

Figure 4. Flowchart for the automatic determination of a sky frame for background subtraction in OSIRIS data.

### 4.3 HIRES extracted spectra

A raw HIRES two-dimensional image contains a number of separate spectral orders spanning across the CCDs. KOA produces the extracted one-dimensional spectra of counts vs. wavelength for each of the orders using the *MAKEE*[6] pipeline. These extracted spectra have been bias-subtracted, flat-fielded, sky-subtracted, and wavelength-calibrated. We also provide products for assessing the quality of the spectral extractions, such as signal-to-noise ratio estimates, sky spectra, flux profile along slit, and centroids of the profiles, all as a function of wavelength.

### 4.4 Calibrating NIRC2 data

NIRC2 calibrated data are available for imaging mode only. Spectroscopic data are currently not offered in calibrated form. When the appropriate calibration files are not available or otherwise cannot be found, no data processing is performed. A description of how calibration files are matched with science files can be found in our calibration association procedure described above.

Each science file is calibrated separately as follows. First, bad pixels are corrected by applying a master bad pixel map and replacing each affected pixel count by the median of the surrounding pixels. Then the data are subtracted by an appropriate dark frame (if one exists) and divided by a flat. We keep a library of master flats and darks produced from median combining a number of images, which are available for several epochs, mainly for broad-band filters and narrow camera (for flats). If available, a master flat and/or dark is used. If not, the median of at least one and a maximum of 10 flats and/or darks found by the calibration association will be used. Finally, if full-frame (i.e., 1024x1024 pixels), the image is corrected for optical distortion following the prescription described on the NIRC2 "dewarp" utility page[7]. Note that for the last step, distortion correction is not performed for sub-arrayed images.

## 5. LESSONS LEARNED

It is useful for KOA to consider its initial experience in creating the original archive for HIRES, and then adding other instruments over the years. Lessons learned, especially from legacy data experience, can help us anticipate what to expect with new instruments, and to take proactive steps to avoid potential problems. Some of the main issues encountered include:

- Definition of keywords

A number of times we found that the nature of the keyword changed for a particular instrument, after we had written the interface control document (ICD)[2]. This required changing the ICD, and often changing the code that handles that specific keyword. To address this issue, for NIRC2 (and subsequent instruments), we performed a complete inspection of all headers, rather than a subset as done for NIRSPEC, in order to bring potential problems to light early in the code writing process.

- Split nights and backup nights

Early on, an algorithm was developed that attempted to properly assign each science file to the proper PI, sometimes a challenge for nights that are split between multiple PIs. However, this was originally done on a file-by-file basis, with no greater context from the entire stream of files taken on that night. An improved algorithm was evolved, and with the addition of observing tools like the PIG, this smoothed out handling nights split between multiple PIs, those with unscheduled ToO programs or with instruments used as backup but not on the telescope schedule.

- Complete keywords to support archiving

Perhaps the best lesson is that in designing the instrument and its data product, it should always be kept in mind that the data are made with the archive in mind. As such, it should be ensured that the full complement of KOA keywords conform to proper standards and are incorporated in the FITS headers well before the instrument produces a single FITS file on the telescope.

**Acknowledgements:** The Keck Observatory Archive (KOA) is a collaboration between the W. M. Keck Observatory (WMKO) and the NASA Exoplanet Science Institute (NExScI). Funding for KOA is provided by the National Aeronautics and Space Administration (NASA). WMKO is operated as a scientific partnership among the California Institute of Technology, the University of California, and NASA. The Observatory was made possible by the generous financial support of the W. M. Keck Foundation. NExScI is sponsored by NASA's Origins Theme and Exoplanet Exploration Program, and operated by the California Institute of Technology in coordination with the Jet Propulsion Laboratory. This material is based in part upon work supported by NASA under Grant and Cooperative agreement No. NNX13AH26A.